\documentclass[aip,jap,reprint,superscriptaddress,amsmath,citeautoscript]{revtex4-1}
\usepackage{graphicx}

\PassOptionsToPackage{breaklinks}{hyperref}
\usepackage{hyperref}
\usepackage[all]{hypcap}
\hypersetup{
	colorlinks = true,
	linkcolor = {blue},
	citecolor = {blue},
	urlcolor = {blue},
}

\begin{document}

\title{Kinematics of slip-induced rotation for uniaxial shock or ramp compression}

\author{P.~G.~Heighway}
\email{patrick.heighway@physics.ox.ac.uk}
\affiliation{Department of Physics, Clarendon Laboratory, University of Oxford, Parks Road, Oxford, OX1 3PU, United Kingdom}
\author{J.~S.~Wark}
\affiliation{Department of Physics, Clarendon Laboratory, University of Oxford, Parks Road, Oxford, OX1 3PU, United Kingdom}

\date{\today}

\begin{abstract}
When a metallic specimen is plastically deformed, its underlying crystal structure must often rotate in order to comply with its macroscopic boundary conditions. There is growing interest within the dynamic compression community in exploiting x-ray diffraction measurements of lattice rotation to infer which combinations of plasticity mechanisms are operative in uniaxially shock- or ramp-compressed crystals, thus informing materials science at the greatest extremes of pressure and strain rate. However, it is not widely appreciated that several of the existing models linking rotation to slip activity are fundamentally inapplicable to a planar compression scenario. We present molecular dynamics simulations of single crystals suffering true uniaxial strain, and show that the Schmid and Taylor analyses used in traditional materials science fail to predict the ensuing lattice rotation. We propose a simple alternative framework based on the elastoplastic decomposition that successfully recovers the observed rotation for these single crystals, and can further be used to identify the operative slip systems and the amount of activity upon them in the idealized cases of single and double slip.
\end{abstract}

\maketitle

\section{\label{sec:introduction} Introduction}

When a crystalline specimen is loaded to stresses exceeding its elastic limit, the plastic deformation that ensues causes its underlying crystal structure to rotate. That is to say that plasticity is usually accompanied by changes in crystallographic texture. Since this observation was first made nearly a century ago\cite{Taylor1923, Mark1923}, tremendous progress has been made on both the modelling and experimental measurement of slip-induced rotation. Modern structural characterization techniques, such as electron backscatter diffraction (EBSD) and three-dimensional x-ray diffraction\cite{Poulsen2001, Lauridsen2001, Poulsen2003} (3DXRD), now make it possible to track the crystallographic orientation of individual grains in quasistatically deforming specimens with microscopic resolution. By comparing such experimental measurements with the predictions of crystal plasticity theory, one can infer the combination of plasticity mechanisms responsible for an observed reorientation of the crystal structure \cite{Margulies2001, Basinski2004, Florando2006, Winther2008, Chen2013, Oddershede2015, Hemery2019}, and thus effectively use texture evolution as a slip-system diagnostic.

It is only very recently that analysis of this kind has been attempted on targets subjected to the dramatic loading conditions of rapid dynamic compression. Modern high-intensity laser-compression platforms allow us to load solid specimens rapidly and reproducibly to the kind of megabar pressure states encountered in planetary interiors. Compression of this kind has historically taken the form of a shock, in which the compression wave is allowed to steepen into a near-discontinuity, resulting in the generation of high-density but also high-entropy material that eventually melts once the shock pressure exceeds a few hundred gigapascals. Increasingly, though, so-called quasi-isentropic compression (QI) techniques are being used that, by the use of temporally shaped laser pulses, multiple shocks, or multilayered targets\cite{McGonegle2020}, allow the sample to be uniaxially compressed in a ramped manner, and thus be kept solid far into the terapascal pressure regime\cite{Bradley2009, Smith2014}. These compression platforms often permit simultaneous use of ultrabright x-ray sources [such as laser-plasma backlighters\cite{Wark1989, Murnane1991, Frederic2016}, synchrotrons\cite{Schoenlein2000, Khan2006, Beaud2007}, or, more recently, hard x-ray free electron lasers\cite{McNeil2010} (XFELs)] to generate time-resolved diffraction images of specimens in the extremely short-lived (i.e.~nanosecond-long) dynamically compressed state. Indeed, it is vital that such measurements are obtained within this nanosecond window, and not \emph{ex post facto}: during the subsequent dynamic release process, it is possible for both compression-induced plasticity\cite{Sliwa2018} and phase transitions\cite{Gorman2019, Gorman2020} to be reversed, meaning recovery experiments can provide only limited information about the properties of the sample in its high-pressure state. Ultrafast x-ray diffraction as a probe of transiently compressed matter is now a mature diagnostic technique, and has facilitated a great number of studies of crystal plasticity\cite{Turneaure2009, Murphy2010, Suggit2012, Milathianaki2013, Comley2013, Wehrenberg2015, Wehrenberg2017, Sliwa2018, Sharma2020} and polymorphic phase transitions\cite{Kalantar2005, Briggs2017, Gorman2018, Coleman2019, Sharma2019, Briggs2019} under extreme loading conditions.

Certain of these studies have focused on the texture evolution of shock-driven specimens. In the seminal work of Wehrenberg \emph{et.~al.}\cite{Wehrenberg2017}, the authors obtained \emph{in situ} diffraction measurements of plasticity-induced rotation in polycrystalline tantalum foils shock-driven to over 200~GPa. The degree of rotation was quantitatively reconciled -- with moderate success -- with the total compression suffered by the samples using a kinematic framework owed to Schmid\cite{Schmid1926, Schmid1935}. In a similar study by Suggit.~\emph{et.~al.}\cite{Suggit2012}, shock-induced lattice rotation of single-crystal copper loaded to around 50~GPa was measured by means of nanosecond Laue diffraction. These results were numerically interpreted\cite{Suggit2012b} using an alternative rotation model attributed to Taylor\cite{Taylor1926, Taylor1927}. These two studies clearly demonstrated that the lattice rotation measurements that have been performed for a century in a traditional materials context can also be performed in the shock-loading regime, and further reflect a growing interest in the dynamic compression community in exploiting texture measurements to study crystal plasticity under the most extreme pressures and deformation rates.

Before advancing these studies, however, it is essential to ensure going forward that the mathematical framework upon which any quantitative texture analysis is based is sound. The two above-mentioned studies appealed to the classic Schmid and Taylor models of slip-induced rotation for a quantitative understanding of the degree of rotation observed. These rotation models were formulated in the context of conventional materials testing scenarios quite unlike that encountered in a shock- or ramp-compression experiment. It is not immediately obvious, then, which of these models (if either) is truly applicable to a uniaxially loaded specimen; indeed Hosford\cite{Hosford1976} and others \cite{Wierzbanowski2011, Wronski2013} have indicated that there is apparently some degree of confusion even in the traditional texture community about when each rotation model is appropriate, and when alternative models might be warranted. The purpose of the present study is to identify and investigate the kinematic description of slip-induced rotation appropriate to uniaxial compression conditions, which does not appear to have been the subject of a dedicated study in the existing literature.

The study is set out as follows. We begin with a brief qualitative discussion of the physics of plasticity-induced texture evolution in Sec.~\ref{sec:texture_evolution}. We then describe in Sec.~\ref{sec:rotation_rules} the Schmid and Taylor rotation rules often used to predict texture evolution in a quasistatic context, and show that they are fundamentally unequipped to treat the uniaxially compressed material encountered in dynamic-compression experiments. We put forward an alternative model based on the elastoplastic decomposition, whose accuracy we verify by testing its predictions against the results of small-scale molecular dynamics simulations. In Sec.~\ref{sec:glide_extraction}, we demonstrate how our simple model may be used to calculate the slip activity of crystals undergoing single and double slip under uniaxial strain conditions. We then discuss those aspects of our model that require further development in Sec.~\ref{sec:discussion}, before concluding in Sec.~\ref{sec:conclusion}.

\section{\label{sec:texture_evolution} Plasticity-induced texture evolution}

\begin{figure}[t]
\includegraphics[scale = 0.92]{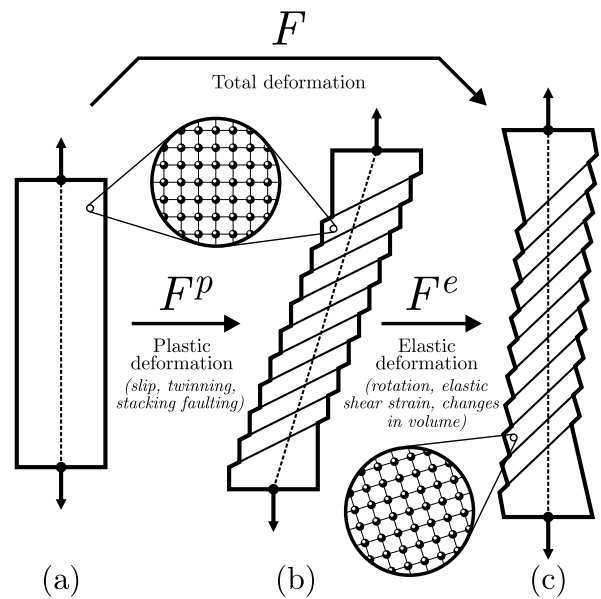}
\caption{\label{fig:tension_test}Rotation attending plastic deformation in a tensile test of a single-crystal sample. A specimen drawn into tension along its long axis suffers plastic deformation $F^p$ via shear motion of its atomic planes. In order to respect the constraints imposed by the total deformation $F$, the specimen must also suffer some degree of elastic deformation, $F^e$. In this instance, the sample must rotate about an axis normal to the slip direction to ensure the ends of the specimen do not become laterally displaced from one another.}
\end{figure}

The essential physics of plasticity-induced texture evolution is frequently illustrated by means of a diagram like Fig.~\ref{fig:tension_test}, which depicts a ductile metallic sample undergoing a simple tensile test. As the specimen is drawn into tension, shear stress accumulates on planes oblique to the loading axis until reaching such a level that sample yields. The specimen subsequently undergoes localized shearing motion on a set of atomic planes (generally assumed to be those under the greatest resolved shear stress at the point of yield) via the glide of dislocations. Depending on the loading conditions and on the material in question, this shearing motion could be realized through full dislocation slip, or via the generation of stacking faults or deformation twins. There is a substantive difference between these two scenarios that will be discussed later (see Sec.~\ref{sec:discussion}), but in either case the specimen will, after shearing, resemble the sample pictured in Fig.~\ref{fig:tension_test}(b). The plastic deformation causes the sample to extend along the direction of glide (which has components both parallel to and perpendicular to the tension axis), but does not, in and of itself, change the crystal structure or orientation of the material between the sheared planes.

Generally, the plastic deformation mediated by the gliding dislocations is incompatible with the \emph{total} deformation to which the sample is subjected -- whatever the loading geometry, there will always be at least one macroscopic constraint upon the sample's orientation, or upon its total dimension in certain directions. To respect these boundary conditions, the specimen must inevitably undergo some degree of elastic deformation. This latter kind of deformation encompasses both lattice rotation and changes to the size or shape of the unit cell. In the case of the tension test pictured in Fig.~\ref{fig:tension_test}, the ends of the sample are held by grips whose motion is confined to the vertical axis. This means that as the sample extends it must also rotate, in order to counteract the transverse displacement caused by the plastic deformation. The sense of the rotation, as illustrated in Fig.~\ref{fig:tension_test}(c), is such as to cause the glide direction to rotate towards the tension axis; if the specimen were instead compressed, the glide direction would rotate away from the loading axis.

While it represents just one contrived loading scenario, the uniaxial tension test neatly illustrates a fundamental point about the deformation of materials in general: the elastic deformation suffered by a specimen as it is loaded (which can often be measured directly in experiment) encodes information about its plastic deformation state (which, in many cases, cannot). In the case of the tension test, the amount of glide on the active slip or twin plane could in principle be inferred from the extent of the lattice rotation, if one also knew the distance through which the grips had moved, i.e.\ the total deformation to which the sample had been subjected. This idea can be mathematically expressed using an \emph{elastoplastic decomposition},  in which the total deformation, expressed by the matrix $F$, is multiplicatively decomposed into a plastic deformation $F^p$ followed by an elastic deformation $F^e$:
	\begin{equation} \label{eq:decomposition}
	F = F^e F^p.
	\end{equation}
If one can claim to know $F$ from the loading conditions and $F^e$ from an experimental measurement, one can deduce the plastic deformation state simply by inverting Eq.~(\ref{eq:decomposition}). From there, one can in principle use the kinematics of crystal plasticity to deduce which combination(s) of plasticity mechanisms would yield the calculated plastic deformation state $F^p$. Eq.~(\ref{eq:decomposition}) thus provides a theoretical framework through which one can garner an understanding of crystal plasticity directly from measurements of a specimen's texture evolution.

For this line of reasoning to work, it is essential that the total deformation gradient $F$, which provides the kinematic link between $F^e$ and $F^p$, is chosen appropriately, so that it accurately reflects the given loading conditions. Different materials testing and processing scenarios (e.g.~tensile testing, channel-die compression, sheet rolling) are described by different experimental boundary conditions, meaning $F$ takes a distinct mathematical form in each case. This in turn alters the manner in which the elements of $F^e$ and $F^p$ and interrelated, and thus the rotation rule one uses to derive slip activity from a given change in crystallographic texture. As stated in Sec.~\ref{sec:introduction}, texture evolution in dynamic compression experiments is currently treated using either the Schmid or Taylor analyses\cite{Suggit2012, Suggit2012b, Wehrenberg2017}, the same treatments one would use to describe the uniaxial tension test. The purpose of the present study is to show that this approach is flawed. While pictures like Fig.~\ref{fig:tension_test} are often used as a visual aid when describing shock-induced rotation, they capture only partially the physics of true uniaxial compression, and therefore give a misleading picture of what is actually taking place at the lattice level behind a planar compression wave. The rotation models and associated formulae that have been `borrowed' from traditional materials science are thus fundamentally inapplicable to shock- and ramp-compression experiments. In the following section, we recap the derivation of the Schmid and Taylor rotation rules, and demonstrate that they fail to predict the texture evolution of a single crystal undergoing uniaxial compression. We then put forward a simple alternative model using an appropriate elastoplastic decomposition, and show that it succeeds where the Schmid and Taylor analyses fail.

\section{\label{sec:rotation_rules} Rotation rules}

Before we begin, we wish to draw the reader's attention to a few technical details about the following derivations. First, we will be using linearized (i.e.~infinitesimal) strain theory throughout. Second, we will restrict our attention to plastic deformation mediated solely by full dislocation slip, rather than deformation twinning or the formation of stacking faults. Third, we will assume implicitly that the modelled specimens, be they mono- or polycrystalline, deform homogeneously. This is to say that the local deformation gradient $F$ at every material point in the sample is identical to the macroscopic deformation gradient, which is in turn determined by the sample's boundary conditions. This is the so-called \emph{full-constraints model}, first put forward by Taylor\cite{Taylor1937}. Doing so gives us a starting point from which to build the simplest model possible, which may be iterated upon in subsequent studies. A discussion of these simplifying assumptions will be reserved for Sec.~\ref{sec:discussion}.

\begin{figure*}
\includegraphics[scale = 0.92]{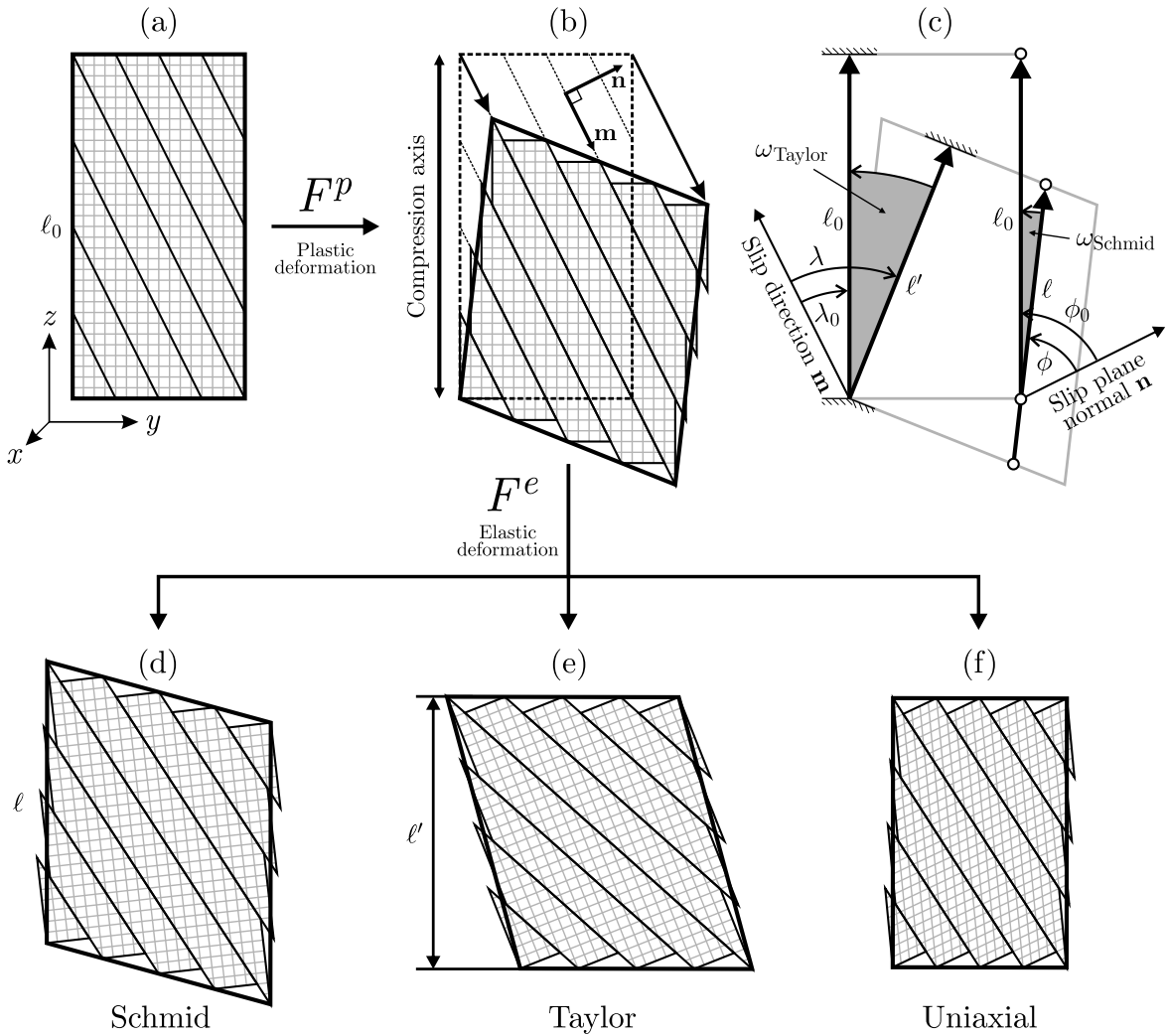}
\caption{\label{fig:rotation_models} Schematic of a material element compressed along the $z$ direction in three different loading scenarios. (a) Element pictured before compression. Diagonal black lines are the intersections of viewing plane with the slip planes about to be activated. The grey grid represents the underlying crystal structure. (b) Element pictured in a hypothetical intermediate state just after plastic deformation via single slip. Vectors $\mathbf{m}$ and $\mathbf{n}$ represent the slip direction and slip plane normal, respectively. (c) Angles and lengths considered in the Schmid and Taylor rotation models. Length $\ell_0$ is the distance between the two ends of the element before compression; lengths $\ell$ and $\ell'$ are the post-plastic-deformation separation of two material points and of two material planes at either end of the sample, respectively. Angles $\phi_0$ and $\phi$ are those made by the slip plane normal with initially vertical line elements before and after compression, respectively. Angles $\lambda_0$ and $\lambda$ are those made by the slip direction with initially vertical plane normals before and after compression, respectively. (d) Element pictured after rotating through angle $\omega_{\text{Schmid}} = \phi_0 - \phi$ so as to preserve the orientation of line elements parallel to $z$. (e) Element pictured after rotating through angle $\omega_{\text{Taylor}} = \lambda - \lambda_0$ so as to preserve the orientation of planar elements normal to $z$. (f) Element pictured after suffering uniaxial strain along $z$, in which both vertical line elements \emph{and} plane normals retain their orientation.}
\end{figure*}

Shown in Fig.~\ref{fig:rotation_models}(a) is a rectangular element of material situated within the bulk of a crystal about to be loaded along the $z$ direction. It will be assumed for now that when the crystal is loaded, only a single slip system becomes active, whose slip direction and slip plane normal will be denoted by $\mathbf{m}$ and $\mathbf{n}$, respectively, with $\mathbf{m} \cdot \mathbf{n} = 0$. For the material pictured in Fig.~\ref{fig:rotation_models}, the loading direction, slip direction, and slip plane normal all lie in the same plane. We note that this simplification is made only to make the underlying physics more transparent, and all of the following theory holds even in the more general case where $(\mathbf{m}\times\mathbf{n})\cdot\mathbf{e}_z \ne 0$.

Upon being compressed along the $z$ direction, the material element responds by shearing on its slip planes, so as to decrease its extent along the loading axis. The mathematical description of this plastic portion of the deformation is well known\cite{Chin1966}: each material point in the element is displaced in a direction parallel to the slip direction $\mathbf{m}$, and by an amount directly proportional to its distance along the slip-plane-normal direction $\mathbf{n}$. This is to say that any line element connecting two material points separated by the vector $\mathbf{r}_0$ is transformed to
	\begin{equation} \label{eq:plastic_remap_line}
	\mathbf{r} = \mathbf{r}_0 + \gamma(\mathbf{r}_0 \cdot \mathbf{n})\mathbf{m},
	\end{equation}
where the glide $\gamma$ expresses the amount of shear motion on the operative slip system. One can show that the transformation above can also be expressed in the form $\mathbf{r} = F^p\mathbf{r}_0$, where the linear operator $F^p$ is the plastic deformation gradient. Explicitly, the elements of $F^p$ read
	\begin{equation} \label{eq:plastic_matrix}
	F^p = \begin{pmatrix}
	1 + \gamma m_x n_x & \gamma m_x n_y & \gamma m_x n_z \\
	\gamma m_y n_x & 1 + \gamma m_y n_y & \gamma m_y n_z \\
	\gamma m_z n_x & \gamma m_z n_y & 1 + \gamma m_z n_z 
	\end{pmatrix},
	\end{equation}
which may be written more succinctly as
	\begin{equation} \label{eq:Fp_otimes}
	F^p = I + \gamma (\mathbf{m} \otimes \mathbf{n}),
	\end{equation}
where $\otimes$ is the outer product operator. As shown in Fig.~\ref{fig:rotation_models}(b), the plastic deformation changes the shape and orientation of the material element, but leaves its underlying crystal structure unaltered. Note also that the plastic deformation conserves the material's volume, which follows immediately from $\text{det}\,F^p=1$.

To accommodate its boundary conditions, the element must also undergo elastic deformation, which is represented by the elastic deformation gradient $F^e$. As noted in the previous section, $F^e$ encodes not only true deformation of the crystal structure (i.e.\ changes to the length of and angles between the lattice vectors), but also local rotation. To disentangle these two effects, $F^e$ is usually further decomposed into a pure elastic strain followed by a rotation (or vice versa) via a \emph{polar decomposition}. The decomposition we opt for here reads
	\begin{equation} \label{eq:polar_decomposition}
	F^e = R^eU^e,
	\end{equation}
where the symmetric \emph{right elastic stretch tensor} $U^e$ accounts for changes to the size and shape of the unit cell, and the rotation matrix $R^e$ expresses the sense and the magnitude of the subsequent lattice reorientation.

When composed, the elastic and plastic parts of the deformation must match the total deformation gradient:
	\begin{equation} \label{eq:decomposition_expanded}
	F = R^eU^e F^p.
	\end{equation}
The value of $F$ is thus the key to linking $F^e$ to $F^p$, and so to constructing the mathematical framework from which one can derive information about a sample's plastic response from the distortion and rotation of its unit cell.

\subsection{Schmid analysis}
The \emph{Schmid analysis} \cite{Schmid1926, Schmid1935} asserts that as the material plastically deforms, it simultaneously rotates in such a way as to preserve the direction of line elements initially aligned with the loading axis, as pictured in Fig.~\ref{fig:rotation_models}(d). This is the behavior exhibited by a long sample that is compressed or stretched along its longest axis, and is allowed to expand or contract freely in the transverse directions (as in the tensile test, for example).

The amount of rotation expected within the Schmid treatment can be derived by applying Eq.~(\ref{eq:plastic_remap_line}) to a vertical line element joining material points at either end of the sample. The projection of this vector onto the slip plane normal $\mathbf{n}$ before and after deformation satisfies the equation
	\begin{subequations}
	\begin{align}
	\mathbf{r} \cdot \mathbf{n} &= \mathbf{r}_0 \cdot \mathbf{n} + \gamma (\mathbf{r}_0 \cdot \mathbf{n})(\mathbf{m} \cdot \mathbf{n}); \\
	\mathbf{r} \cdot \mathbf{n} &= \mathbf{r}_0 \cdot \mathbf{n}.
	\end{align}
	\end{subequations}
Therefore if the distance between the ends of the sample before and after the plastic stage of deformation are $\ell_0$ and $\ell$, respectively, it follows that
	\begin{equation} \label{eq:schmid_ratio}
	\ell\cos\phi = \ell_0\cos\phi_0,
	\end{equation}
where $\phi_0$ and $\phi$ are the angles made by the slip plane normal with the compression direction before and after compression, respectively. As illustrated in the right-hand side of Fig.~\ref{fig:rotation_models}(c), to preserve the direction of its vertical axis, the specimen must rotate through angle $\omega_{\text{Schmid}} = \phi_0 - \phi$. It then follows from Eq.~(\ref{eq:schmid_ratio}) that
	\begin{equation} \label{eq:schmid_rotation}
	\omega_{\text{Schmid}} = \phi_0 - \arccos\left(\frac{\cos\phi_0}{\ell/\ell_0}\right).
	\end{equation}
In the context of the experimental arrangement for which the Schmid analysis was intended, the quantity $\ell/\ell_0$ coincides with the total engineering strain along the loading direction, which is readily measured in experiment.

\subsection{Taylor analysis}
The \emph{Taylor analysis} \cite{Taylor1926, Taylor1927} is predicated on the idea that it is not line elements parallel to the loading axis that retain their orientation during deformation, but planes normal to $z$ that do so. This is the behavior one would expect of a short, wide sample loaded along its shortest axis -- in this instance, the boundary conditions `prioritize' keeping the faces of the specimen in contact with those of the plates loading it.

To derive the rotation expected within the Taylor treatment, one can exploit the fact that there exists an equation analogous to Eq.~(\ref{eq:plastic_remap_line}) for the transformation of planar material elements. Under the action of a linear operator $F$, the vector $\mathbf{s}_0$ that is normal to a particular set of material planes and whose magnitude is \emph{inversely} proportional to their separation transforms according to $\mathbf{s}_0 \to [F^T]^{-1}\mathbf{s}_0$. It may be shown using Eq.~(\ref{eq:plastic_matrix}) that for the present case of plastic deformation,
	\begin{equation}
	[(F^p)^T]^{-1} = I - \gamma (\mathbf{n}\otimes\mathbf{m}).
	\end{equation}
Hence, for the purposes of transforming material planes, one can exchange the roles of $\mathbf{m}$ and $\mathbf{n}$, and invert the sign of the glide $\gamma$. This implies that
	\begin{equation} \label{eq:plastic_remap_plane}
	\mathbf{s} = \mathbf{s}_0 - \gamma (\mathbf{s}_0 \cdot \mathbf{m})\mathbf{n}.
	\end{equation}
We can now apply Eq.~(\ref{eq:plastic_remap_plane}) to the vector normal to the planes at either end of the sample. The projection of this vector onto the slip direction $\mathbf{m}$ before and after deformation satisfies the equation
	\begin{subequations}
	\begin{align}
	\mathbf{s} \cdot \mathbf{m} &= \mathbf{s}_0 \cdot \mathbf{m} - \gamma (\mathbf{s}_0 \cdot \mathbf{m})(\mathbf{n} \cdot \mathbf{m}) \\
	\mathbf{s} \cdot \mathbf{m} &= \mathbf{s}_0 \cdot \mathbf{m}
	\end{align}
	\end{subequations}
Therefore if the distance between the ends of the sample before and after deformation are $\ell_0$ and $\ell'$ respectively, it follows that
	\begin{equation} \label{eq:taylor_ratio}
	\frac{\cos\lambda}{\ell'} = \frac{\cos\lambda_0}{\ell_0}.
	\end{equation}
where $\lambda_0$ and $\lambda$ are the angles made by the slip direction with the compression direction before and after compression, respectively. As illustrated in the left-hand side of Fig.~\ref{fig:rotation_models}(c), to preserve the orientation of the vertical faces, the specimen must rotate through angle $\omega_{\text{Taylor}} = \lambda - \lambda_0$. It follows from Eq.~(\ref{eq:taylor_ratio}) that
	\begin{equation} \label{eq:taylor_rotation}
	\omega_{\text{Taylor}} = \arccos\left(\cos\lambda_0 \frac{\ell'}{\ell_0}\right) - \lambda_0,
	\end{equation}
where $\ell'/\ell_0$ is, again, the engineering strain.

\subsection{Uniaxial analysis\label{sec:uniaxial_analysis}}
The uniaxial compression conditions encountered in a planar shock- or ramp-loading scenario differ fundamentally from those treated by the Schmid and Taylor analyses. The mathematical expression of true uniaxial strain reads
	\begin{equation}
	F =
	\begin{pmatrix}
	1 & 0 & 0 \\ 0 & 1 & 0 \\ 0 & 0 & v
	\end{pmatrix},
	\end{equation}
where $v = V/V_0$ is the ratio of the specimen's volumes before and after compression. This form for the total deformation gradient reflects the assumption that material within the bulk of the sample is everywhere prevented from expanding or contracting in the directions perpendicular the compression direction by the material surrounding it -- this is termed \emph{lateral confinement}. In fact, the bulk material is assumed to suffer no overall distortion or rotation at all, but simply contracts along the compression direction. For such material, it may be said that both the Schmid and Taylor treatments are in some sense true: vertical line elements retain their orientation, as do material planes normal to the loading axis. However, these conditions cannot be satisfied \emph{simultaneously} if the underlying crystal structure is allowed only to rotate. For the material to accommodate the uniaxial strain boundary conditions, it must also suffer some degree of elastic pure shear strain. This is to say that the unit cell not only rotates, but also changes shape. This conclusion can be reached either by close inspection of Fig.~\ref{fig:rotation_models}(f), or by examination of the components of the elastoplastic decomposition [Eq.~(\ref{eq:decomposition_expanded})], which, for this particular instance of single slip in the $yz$ plane, read
\begin{widetext}
\begin{equation} \label{eq:decomposition_full}
	\underbrace{\begin{pmatrix}
	1 & 0 & 0 \\
	0 & 1 & 0 \\
	0 & 0 & v
	\end{pmatrix}}_{F}
	=
	\underbrace{\begin{pmatrix}
	1 & 0 & 0 \\
	0 & \cos\omega & -\sin\omega \\
	0 & \sin\omega & \cos\omega
	\end{pmatrix}}_{R^e}
	\underbrace{\begin{pmatrix}
	1 & 0 & 0 \\
	0 & U_{yy}^e & U_{yz}^e \\
	0 & U_{yz}^e & U_{zz}^e
	\end{pmatrix}}_{U^e}
	\underbrace{\begin{pmatrix}
	0 & 0 &  0 \\
	0 & 1 + \gamma m_y n_y & \gamma m_y n_z \\
	0 & \gamma m_z n_y & 1 + \gamma m_z n_z 
	\end{pmatrix}}_{F^p},
	\end{equation}
\end{widetext}
where $\omega$ measures the lattice rotation about the $x$ axis. One will find that the equation above cannot be solved if one only permits the crystal structure to rotate -- one needs the additional degrees of freedom offered by the elastic strains $U_{ij}^e$ to find an internally consistent solution that fulfils the uniaxial strain boundary conditions.

To derive an equation for the crystal rotation analogous to Eqs.~(\ref{eq:schmid_rotation}) and (\ref{eq:taylor_rotation}) is a simple matter of inverting the equation above to solve for the rotation matrix $R^e$, and then using two of its elements to eliminate the unknown glide~$\gamma$ (as we will show shortly). However, it should be borne in mind that for these uniaxial strain conditions, the rotation $\omega$ does not hold quite the same meaning that it did in the Schmid and Taylor pictures, because it no longer tells the whole story -- the crystallographic texture now depends on both the rotation matrix $R^e$ \emph{and} on the nonzero deviatoric components of the elastic stretch tensor $U^e$.  This is to say that there is now a real difference between rotation in the `technical' sense (that is, as the antisymmetric part of the elastic deformation gradient, $R^e$) and rotation in the `generic' sense of anything that changes the orientation of a crystal's atomic planes. With this in mind, we now go on to compare the rotation predictions from the Schmid, Taylor, and uniaxial analyses with the results of classical molecular dynamics (MD) simulations, which have become indispensable tools for modelling dynamically loaded matter over the length- and timescales pertinent to shock and ramp compression\cite{Holian1998, Germann2000, Bringa2006, Ravelo2013, Gunkelmann2015, Higginbotham2016, Tang2017}.

\subsection{\label{sec:verification} Verification of kinematics via small-scale MD}
To verify that the kinematics expressed by the elastoplastic decomposition [Eq.~(\ref{eq:decomposition_full})] are sound, we performed small-scale MD simulations of monocrystalline tantalum uniaxially compressed so as to induce single slip. We used a form of nearest-neighbor analysis to calculate the average rotation of the unit cell that ensued, and compared its value with predictions based on both the traditional Schmid and Taylor analyses, and on the full elastoplastic decomposition for uniaxial strain. Predictions obtained using the former approaches differ from the true rotation by approximately 20\%, while the latter method gives agreement to within a few percent. Before discussing these results, we will first describe the setup and characterization of the simulations, all of which were executed using the open-source code \textsc{lammps} \cite{Plimpton1995}.

\begin{figure}[t]
\includegraphics{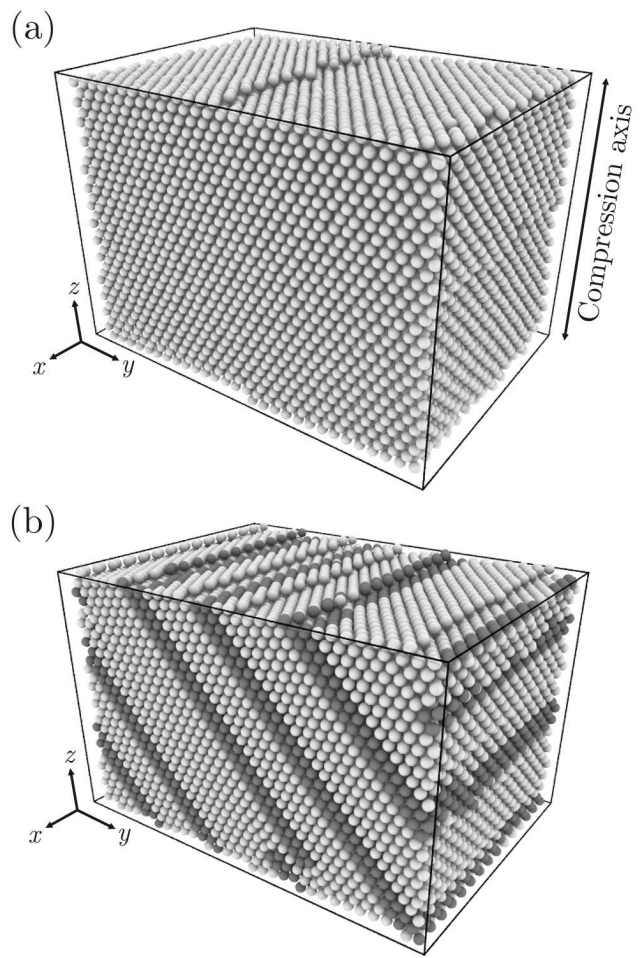}
\caption{\label{fig:quasistatic_2D_schematic} Small-scale molecular dynamics simulation of a block of bcc tantalum uniaxially compressed so as to induce single slip, visualized using Ovito\cite{Stukowski2010}. (a) The sample before compression, spanning $70.2\times99.3\times70.3$~\AA\textsuperscript{3} along the coordinate axes and subjected to periodic boundary conditions on every one of its faces. The crystallographic directions initially aligned with $x$, $y$, and $z$ are $[1\,0\,\bar{1}]$, $[\bar{1}\,30\,\bar{1}]$, and $[15\,1\,15]$, respectively. (b) The sample after compression by 12\% along $z$. Darker atoms are those whose original neighbors have been displaced by vector $\pm[1\bar{1}1]$, which are found on the $(121)$ planes.}
\end{figure}

Fig.~\ref{fig:quasistatic_2D_schematic}(a) pictures a typical crystal simulated here before compression. It consists of 27\,060 tantalum atoms arranged in a body-centered-cubic (bcc) configuration with a cubic lattice constant of $a_0 = 3.309$~\AA, found within a simulation box spanning $70.2\times99.3\times70.3$~\AA\textsuperscript{3} along the coordinate axes. The peculiar aspect ratio of the box results from the low-symmetry choice of crystallographic orientation (which will be explained shortly) combined with the need for periodic boundary conditions, which imitate the presence of surrounding material that would confine matter deep in the bulk of a rapidly loaded sample. A slightly elevated initial temperature of 500~K for the crystal is chosen so as to encourage full dislocation slip over deformation twinning. Interactions between neighboring atoms are modelled using the embedded-atom-method (EAM) potential Ta2 developed by Ravelo \emph{et.\ al.}\ \cite{Ravelo2013}, which was specifically developed for high-pressure applications of the kind here.

The orientation of the crystal is chosen such that the compression axis lies close to, but not quite parallel to, the $[101]$ direction. In a bcc crystal compressed along $[101]$ exactly, there exist just two symmetrically equivalent slip systems -- $[1\bar{1}1](121)$ and $[111](1\bar{2}1)$ -- that both experience the greatest resolved shear stress. By tilting the $[101]$ direction slightly away from the compression axis, one can break this symmetry, and thus encourage the crystal to `choose' one of these two slip systems over the other when it yields. For this reason, we orient the crystal such that $[1\,0\,\bar{1}]$, $[\bar{1}\,30\,\bar{1}]$, and $[15\,1\,15]$ are aligned with the $x$, $y$, and $z$ directions, respectively. This orientation results in the $[101]$ direction being pre-tilted $2.7^\circ$ away from~$z$. This, as we shall see, is enough to ensure that only one of the above-mentioned primary slip systems is activated. We note in passing that the small dimensions of the crystal also tend to suppress plasticity \cite{Holian1998, Kimminau2010}, and thus further encourage single slip.

The crystal is first briefly thermalized for one picosecond under constant-NVE conditions to bring it to the desired temperature. It is then uniaxially and uniformly compressed along $z$ over the course of 5~ps (via a simple rescaling of its atomic coordinates) until it reaches 88\% of its original volume. The pressure in the crystal at this point is just under 33~GPa. The now-metastable crystal is then held at constant volume for 20~ps, during which time it yields and deforms plastically until the resolved shear stress acting on the active slip system drops below the flow stress. Once the crystal has equilibrated, its temperature is then reduced to approximately 100~K over the course of 20~ps with a Langevin thermostat, to damp out thermal fluctuations that would otherwise broaden the distribution of local strain and rotation states. The final atomistic configuration is pictured in Fig.~\ref{fig:quasistatic_2D_schematic}(b); this visualization and all others herein are from Ovito\cite{Stukowski2010}.

To confirm that the crystal does indeed slip on only a single slip system in response to the uniaxial compression, we employ a variant of slip vector analysis (SVA) \cite{Zimmerman2001}. This technique, whose implementation is described in full in Ref.~[\onlinecite{Heighway2019}], categorizes each atom according to the slip event(s) in which it has participated by examining the displacements of its nearest neighbors from their original positions. In this instance, SVA reveals that a subset of the atoms have had some of their neighbors displaced by a vector consistent with $\pm[1\bar{1}1]$. When these atoms are shaded distinctly, as in Fig.~\ref{fig:quasistatic_2D_schematic}(b), one observes that they form planes of the type $(121)$, and none other. This allows us to identify $[1\bar{1}1](121)$ as the sole active slip system.  We hence deduce the appropriate components of the slip direction and slip plane normal in the simulation box basis, which read $\mathbf{m} = (0,0.6152,-0.7884)$ and $\mathbf{n} = (0,0.7884,0.6152)$, respectively.

To calculate the elastic deformation gradient of the crystal, we use a form of nearest-neighbor analysis that characterizes each atom's unit cell. Again, the algorithm is described in full in Ref.~[\onlinecite{Heighway2019}], but we will recap the essential details here. Atoms in noncrystalline environments (i.e.\ those located near crystal defects, for which an elastic deformation gradient cannot be uniquely defined) are first excluded from the computation with an adaptive common neighbor analysis (a-CNA) \cite{Stukowski2012b} prefilter. The algorithm then takes each nondefective atom in turn, and pairs each of its neighbors with a nearest-neighbor vector taken from a `template' bcc structure representing the original shape and orientation of the unit cell. The linear operator $F^e$ that maps the original structure onto the current structure is then calculated, and subsequently decomposed into the stretch and rotation tensors $U^e$ and $R^e$, per Eq.~(\ref{eq:polar_decomposition}). These per-atom tensors may then be binned in order to calculate the distribution of elastic strain and rotation states present in the crystal.

\begin{figure}[t]
\includegraphics{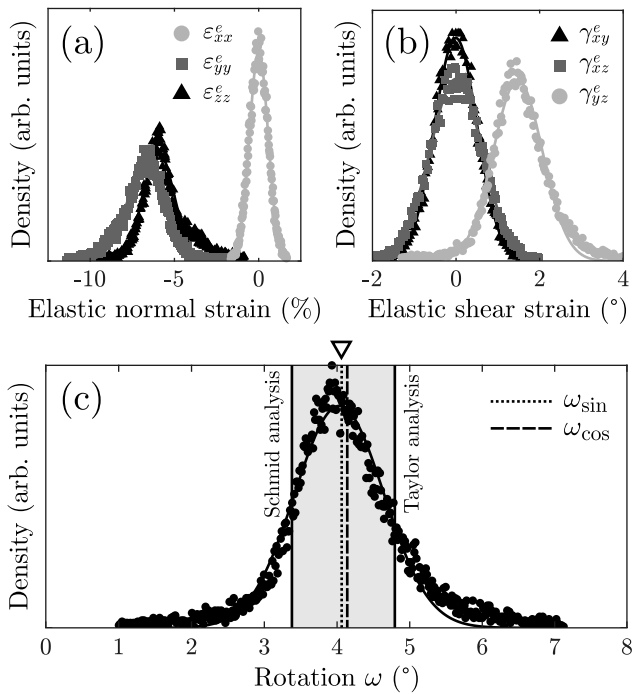}
\caption{\label{fig:quasistatic_compression_validation} Distributions of the elements of the tensors $U^e$ and $R^e$ for a uniaxially compressed tantalum crystal undergoing single slip on its $[1\bar{1}1](121)$ system, situated in the $yz$ plane. (a)~On-diagonal elements of the elastic stretch tensor $U^e$, expressed as engineering strains $\varepsilon^e_{ii} = U_{ii}^e - 1$. (b)~Off-diagonal elements of $U^e$, expressed as engineering shear strains $\gamma_{ij}^e = 2U_{ij}^e$. (c)~Rotation about $x$, calculated via $\omega = \arcsin(R_{zy}^e)$. Also shown are predictions of the mean rotation based on the Schmid and Taylor analyses, and from the full elastoplastic decomposition. The mean of the rotation distribution is indicated by the hollow triangle.}
\end{figure}

Fig.~\ref{fig:quasistatic_compression_validation} shows the distributions of elements of the tensors $U^e$ and $R^e$  for the plastically strained crystal. Several features are immediately apparent. First, Fig.~\ref{fig:quasistatic_compression_validation}(a) shows that the normal elastic strains in the $yz$ plane are approximately equal, owing to the plastic deformation taking place in that plane. Elastic strain component $\varepsilon_{xx}^e$, meanwhile, is still centered on zero, which is to be expected given that the plasticity causes no motion in the $x$ direction. Second, we see from Fig.~\ref{fig:quasistatic_compression_validation}(b) that there is indeed a nonzero amount of elastic pure shear strain in the $yz$ plane. This, as discussed in Sec.~\ref{sec:uniaxial_analysis}, follows inevitably from the uniaxial strain boundary conditions. The degree of shear strain is such that the angle between lattice vectors initially aligned with $y$ and $z$ would change by just under $2^\circ$. Third, we observe in Fig.~\ref{fig:quasistatic_compression_validation}(c) that the crystal does indeed rotate about the $x$ axis (by approximately $4^\circ$) such that the slip direction rotates away from $z$. The physics expressed by the elastoplastic decomposition therefore appears to be borne out by the simulations.

To verify that Eq.~(\ref{eq:decomposition_full}) is also quantitatively correct, we can calculate the rotation $\omega$ that it predicts given a particular state of elastic strain. This can be done by inverting Eq.~(\ref{eq:decomposition_full}) such that $R^e$ is isolated, and then eliminating $\gamma$ between equations for two of its components. By combining the on-diagonal components of $R^e$, one obtains the following expression for the rotation:
	\begin{equation} \label{eq:omegacos}
	\cos\omega_{\text{cos}} = \frac{W_1 U_{yy}^e - W_2 U_{zz}^e}{W_1 - W_2 v},
	\end{equation}
where
	\begin{subequations}
	\begin{align} \label{eq:omegacos_components}
	W_1 &= U_{yz}^e m_y n_z + U_{zz}^e m_z n_z, \\
	W_2 &= U_{yz}^e m_z n_y + U_{yy}^e m_y n_y.
	\end{align}
	\end{subequations}
If instead one combines the off-diagonal components, one arrives at
	\begin{equation} \label{eq:omegasin}
	\sin\omega_{\text{sin}} = -U_{yz}\frac{W_3 - W_4}{W_3 + W_4 v},
	\end{equation}
where
	\begin{subequations} \label{eq:omegasin_components}
	\begin{align}
	W_3 &= U_{yy}^e m_y n_z + U_{yz}^e m_z n_z, \\
	W_4 &= U_{zz}^e m_z n_y + U_{yz}^e m_y n_y.
	\end{align}
	\end{subequations}
If the kinematics are correct, $\omega_{\text{cos}}$ and $\omega_{\text{sin}}$ should of course be identical to within thermal noise. For comparison, we can also derive rotation predictions based on the Schmid and Taylor analyses. For material exhibiting pronounced elastic-plastic behavior under uniaxial strain conditions, the combined quantity $\ell/\ell_0$ appearing in Eqs.~(\ref{eq:schmid_rotation}) and (\ref{eq:taylor_rotation}) can be shown to be equal to
	\begin{equation}
	\frac{\ell}{\ell_0} = \sqrt{(F_{yy}^e)^2 + (F_{yz}^e)^2},
	\end{equation}
where the elements of $F^e$ can be derived from those of $U^e$ and $R^e$ via Eq.~(\ref{eq:polar_decomposition}). As noted previously, the Schmid and Taylor treatments are, strictly speaking, fundamentally inapplicable to the uniaxial strain scenario, but it is informative to see the degree of rotation they predict when they are applied in good faith.

In Fig.~\ref{fig:quasistatic_compression_validation}(c), we compare rotations predicted by the different rotation models with the true rotation distribution. As anticipated, neither the Schmid nor the Taylor analyses predict quite the right rotation. They do, however, bound the correct answer. The intuition for this follows from Fig.~\ref{fig:rotation_models}(b): the sense of the elastic shear strain must be such as to reduce the obtuse angle between the edges of the material element, meaning the element must rotate further than the Schmid analysis would suggest in order to keep its vertical edge aligned with $z$, but somewhat less than in the Taylor picture to preserve the orientation of its upper and lower faces.
Meanwhile, the predictions from the full elastoplastic decomposition, $\omega_{\text{cos}}$ and $\omega_{\text{sin}}$, agree with the mean rotation to within better than 2\%. Note also that $\omega_{\text{cos}}$ and $\omega_{\text{sin}}$ agree with one another to within a few percent. We have repeated these small-scale compression simulations several times with different thermal seeds, and verified that this level of agreement is reproducible. From these results, we conclude that the kinematic framework expressed by Eq.~(\ref{eq:decomposition_full}) is both internally consistent and accurate for the case of single slip under uniaxial strain conditions.

\section{\label{sec:glide_extraction} Glide extraction}

Now that we have verified our simple model of rotation under uniaxial strain conditions, we can attempt to use it for its intended purpose: to calculate the amount of glide on the active slip system(s) from the crystal's state of elastic deformation. We will first revisit the case of single slip, before moving onto an idealized case of equal slip on two complementary systems in Sec.~\ref{sec:double_slip}.

\subsection{\label{sec:single_slip} Single slip}

In the previous section, we took several measures to guarantee that just one slip system became active under uniaxial compression. These measures included limiting the dimensions of the crystal, and selecting a very particular crystallographic orientation and compression ratio $v$. While it happens to succeed in bringing about the desired plastic deformation mode in bcc tantalum compressed along $\sim[101]$, this scheme does not readily generalize to arbitrary orientations and compressions. For this reason, we shall hereon no longer rely on homogeneous nucleation to generate the dislocations we require, but instead adopt the technique of deliberately injecting one or more dislocations into the crystal before compression begins. This is a well-established technique in molecular dynamics (often used to study dislocation mobility\cite{Chang2002,Marian2004,Marian2006,Barton2011,Maresca2018}) that allows us to exercise far greater control over the manner in which the crystal plastically deforms. The first case we shall examine is single slip in a somewhat-larger-scale tantalum crystal, this time with its $[001]$ direction aligned with $z$, which was constructed as follows.

\begin{figure}
\includegraphics{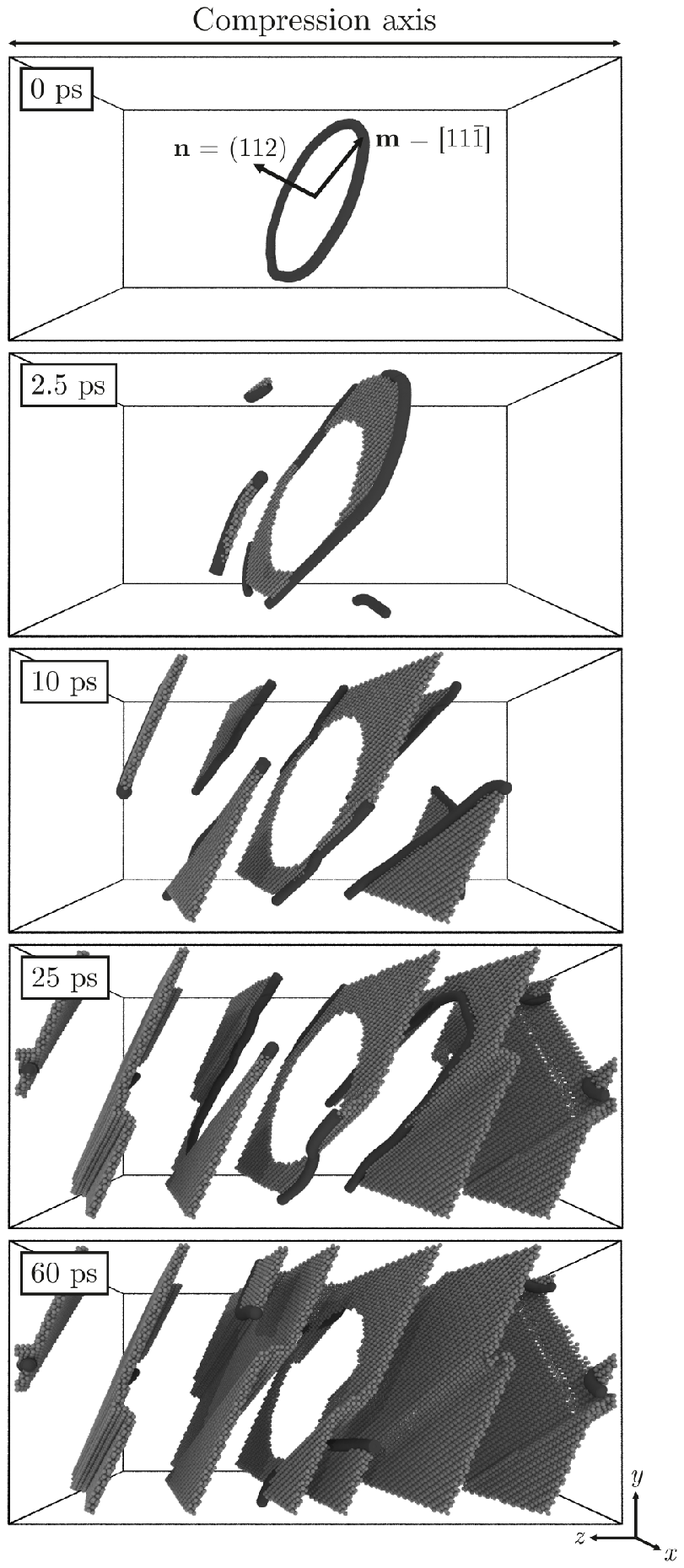}
\caption{\label{fig:single_slip_slideshow} Evolution of a fully periodic block of tantalum containing a pre-existing dislocation loop of the type $[11\bar{1}](112)$ when compressed by 10\% along $[001]$. The crystal initially has dimensions of $16.5\times16.5\times29.8$\,nm\textsuperscript{3}, and has its $\langle100\rangle$ crystallographic directions aligned with the simulation cell axes. Shown are both the dislocation loop, which was characterized using the dislocation extraction algorithm (DXA) \cite{Stukowski2010b,Stukowski2012}, and the slipped atoms left in its wake as it propagates, which were identified using slip vector analysis (SVA)\cite{Zimmerman2001}. Atoms slipped during the injection of the loop itself have been omitted. Visualizations were performed using Ovito \cite{Stukowski2010}. }
\end{figure}

A fully periodic block of defect-free bcc tantalum with dimensions of $16.5\times16.5\times29.8$\,nm\textsuperscript{3} is first built with its $\langle100\rangle$ directions collinear with the coordinate axes and with an initial temperature of 300~K. Before any time integration takes place, two 16-nm-wide circular layers of atoms situated on adjacent $(112)$ planes are identified and `frozen', such that they no longer respond to interatomic forces. Following the prescription of Verschueren \emph{et. al.}~\cite{Verschueren2017}, these adjacent planes are then symmetrically displaced from one another through vectors $\pm\frac{1}{4}[11\bar{1}]$ over the course of 5~ps, while the remainder of the crystal evolves under an NVE scheme as usual. Once the planes are fully separated, the entire crystal is simulated for a further 5~ps under NVE conditions with an added Langevin thermostat, whose purpose is to damp out the strain waves emitted by the nascent dislocation. A visualization of the relaxed $\frac{1}{2}[11\bar{1}]$ dislocation loop obtained immediately before compression begins is shown at the top of Fig.~\ref{fig:single_slip_slideshow}.

The crystal is then instantaneously compressed by 10\% along its $[001]$ axis. The attendant shear stress is insufficient to create any new dislocations, but is great enough to drive the existing dislocation loop, and thus cause slip on the $[11\bar{1}](112)$ system. The progression of the dislocation loop and the slipped atoms left in its wake are illustrated in Fig.~\ref{fig:single_slip_slideshow}. We observe that during the first few picoseconds of the simulation, the dislocation loop grows principally along a direction collinear with its Burgers vector $\frac{1}{2}[11\bar{1}]$. This is consistent with the general trend for edgelike dislocations to be considerably more mobile than their screwlike counterparts in bcc crystals \cite{HullandBacon2011, Queyreau2011, Chaussidon2006, Monnet2009, Chen2020}. During the following 30~ps, the edgelike segments of the dislocation loop rapidly propagate and traverse the periodic boundaries several times, all the while relieving the very shear stress driving them. After approximately 60~ps, an equilibrium state is reached in which slip planes pervade the crystal uniformly, and the shear stress acting on the dislocation has dropped to such a level that further plastic flow is inhibited.

We should note that crystal does not in fact undergo \emph{perfect} single slip -- close inspection of Fig.~\ref{fig:single_slip_slideshow} reveals several `wrinkles' in the slip planes. These are formed wherever screwlike segments of the dislocation loop undergo cross-slip, i.e.~temporarily transfer to a slip plane other than $(112)$. The relatively low temperature and shear stress state of the crystal means the degree of cross-slip is low enough that the results of our analysis, which assumes perfect single slip, are not adversely affected. In a sample dynamically compressed to much greater stresses, however, the dislocations may not be so forgiving, and it might be necessary to formulate a more general model allowing for glide on unanticipated slip systems -- this is the subject of Sec.~\ref{sec:taylor_ambiguity}.

\begin{figure}
\includegraphics{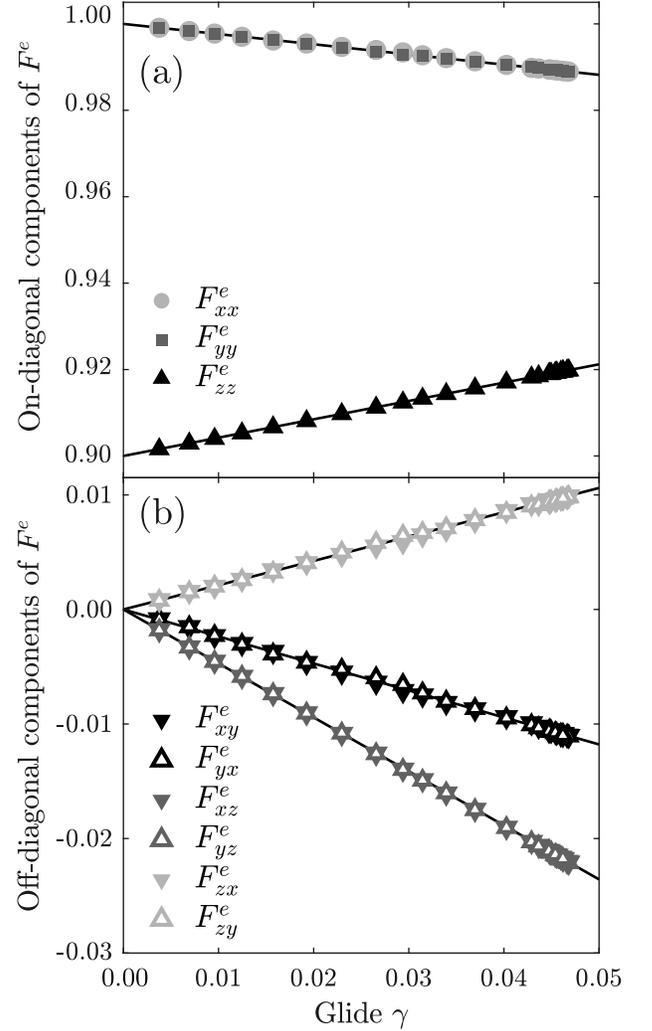}
\caption{\label{fig:loci_single} Evolution of the elements of the average elastic deformation gradient $F^e$ for a tantalum crystal uniaxially compressed by 10\% along $[001]$, deforming predominantly on its $[11\bar{1}](112)$ slip system. Elements of $F^e$ are plotted as a function of the instantaneous glide $\gamma$ fitted from the elastoplastic decomposition. Solid lines are the loci to which the elements of $F^e$ should theoretically adhere. (a)~Evolution of the on-diagonal elements. (b)~Evolution of the off-diagonal elements.}
\end{figure}

If we neglect the relatively small amount of `contaminant' cross-slip, we may appeal once again to the single-slip form of the elastoplastic decomposition for a kinematic description of the plastic deformation process. Rather than using this framework to look for internal consistency between the elastic strains and rotation, as in the previous section, we shall instead use it to connect directly the elastic and plastic deformation states. By simple inversion of Eq.~(\ref{eq:decomposition}), one finds the elastic deformation state for a given state of plastic strain reads $F^e = FF^{p-1}$. For the present case of $\mathbf{m} = [11\bar{1}]$, $\mathbf{n} = (112)$, one may show that, explicitly,
	\begin{equation} \label{eq:locus_single}
	F^e(\gamma) = 
	\begin{pmatrix}
	1 - \frac{\sqrt{2}}{6}\gamma & -\frac{\sqrt{2}}{6}\gamma & -\frac{\sqrt{2}}{3}\gamma \\
	-\frac{\sqrt{2}}{6}\gamma & 1 - \frac{\sqrt{2}}{6}\gamma & -\frac{\sqrt{2}}{3}\gamma \\
	\frac{\sqrt{2}}{6}\gamma v & \frac{\sqrt{2}}{6}\gamma v & \left(1 + \frac{\sqrt{2}}{3}\gamma\right)v,
	\end{pmatrix}
	\end{equation}
where in this instance $v = 0.90$. The equation above essentially parametrizes the locus of elastic deformation states accessible to the crystal, given its single degree of freedom $\gamma$. Our objective here is to invert the equation above and solve for $\gamma$ (in a least-squares sense) using known values of $F^e_{ij}$, and in so doing infer the degree of activity on the operative slip plane from the size, shape, and orientation of the crystal's unit cell.

\begin{figure}
\includegraphics{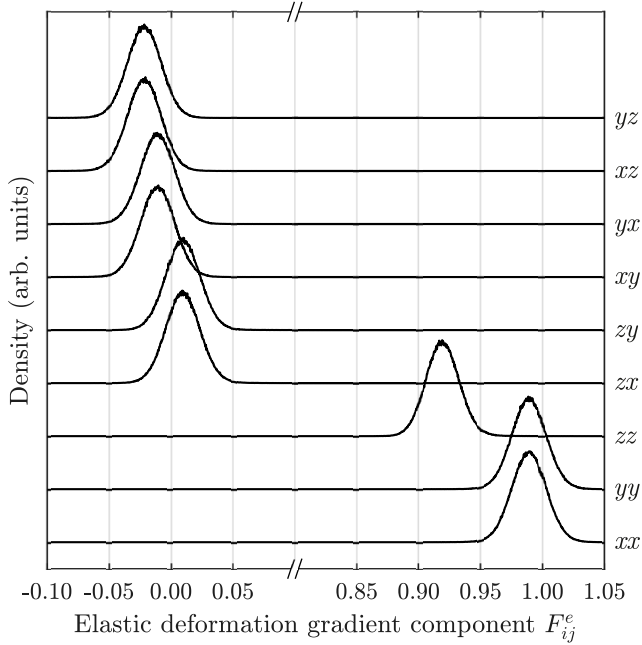}
\caption{\label{fig:distributions_waterfall_single} Distributions of the elements of $F^e$ for a tantalum crystal uniaxially compressed by 10\% along $[001]$, deforming predominantly on its $[11\bar{1}](112)$ slip system. Distributions have been vertically offset for clarity, and ordered in such a way as to make the symmetries of $F^e$ clearer. Distributions were taken from the latest simulation timestep at $t = 60$~ps.}
\end{figure}

Shown in Fig.~\ref{fig:loci_single} are the nine elements of the elastic deformation gradient $F^e$ calculated at 2.5~ps intervals, where the abscissa is the fitted glide obtained from Eq.~(\ref{eq:locus_single}) at each instant of time. These data points are plotted over the theoretical variations of $F^e_{ij}$ with $\gamma$ expected from Eq.~(\ref{eq:locus_single}). We observe that the observed elastic deformation state does indeed adhere very closely to the theoretical locus of states, from the initial unrelaxed state in which $F^e \approx F = \text{diag}(1,1,v)$, to the final relaxed state of relatively low elastic shear strain. All of the symmetries expected from Eq.~(\ref{eq:locus_single}), namely
\begin{subequations}
	\begin{align}
	\langle F_{xx}^e\rangle &= \langle F_{yy}^e\rangle, \\
	\langle F_{xy}^e\rangle &= \langle F_{yx}^e\rangle, \\
	\langle F_{xz}^e\rangle &= \langle F_{yz}^e\rangle, \\
	\langle F_{zx}^e\rangle &= \langle F_{zy}^e\rangle.
	\end{align}
	\end{subequations}
are correctly borne out by the simulations. This is made particularly apparent when one inspects the full distributions of the elements of $F^e$, an example of which is provided by Fig.~\ref{fig:distributions_waterfall_single}. We see that the distributions of $F_{xx}^e$ and $F_{yy}^e$, for instance, are practically identical to within thermal noise, just as expected.

\begin{figure}
\includegraphics{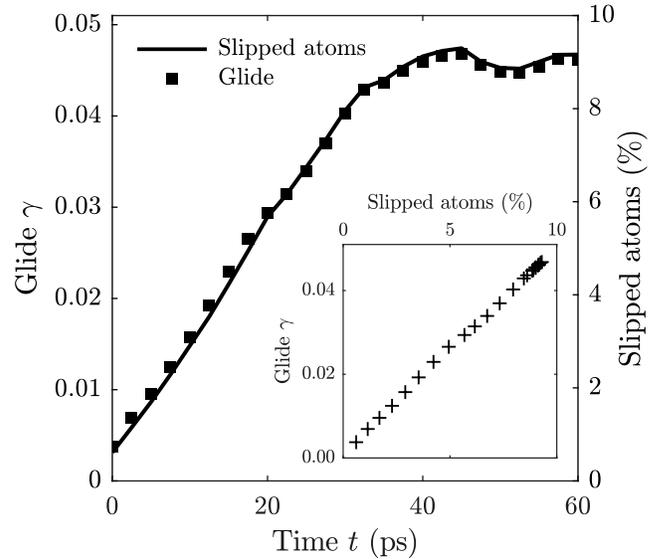}
\caption{\label{fig:glide_vs_slipped_atoms_single} Measurements of slip activity in a tantalum crystal uniaxially compressed along $[001]$, deforming predominantly on its $[11\bar{1}](112)$ slip system. Discrete points show the glide derived from the elastic deformation gradient, while the solid curve shows the total number of slipped atoms. The inset illustrates the correlation between these two independent measures of slip activity.}
\end{figure}

We can now compare the glide extracted from the elastoplastic decomposition with the `true' amount of glide calculated from the total area of faulted material. The latter measurement can be performed simply by counting up the number of slipped atoms detected using SVA, including those resulting from the initial injection of the dislocation. We show in Fig.~\ref{fig:glide_vs_slipped_atoms_single} the fitted glide as a function of time, $\gamma(t)$, alongside the total number of slipped atoms, expressed as a percentage of the total population of the crystal. The two independently calculated measurements of slip activity show almost identical trends. We can therefore assert that the scheme works as intended: given the average elastic deformation state of a uniaxially compressed single crystal, we can calculate the total amount of slip accrued on a single, predetermined slip system.

\subsection{\label{sec:double_slip} Double slip}

While it makes for a useful test case, single slip is arguably of limited relevance to dynamic uniaxial compression. This is because under the conditions of lateral confinement, a single slip system can only relieve certain components of the shear stress. For the [101] crystal studied in Sec.~\ref{sec:verification}, for instance, we observed that single slip left the atomic spacing normal to $\mathbf{m}$ and $\mathbf{n}$ completely unchanged. We note in passing that there exists some evidence that grains in a shock-loaded nanocrystalline specimen might be able to relieve remaining shear stress components elastically, by deforming cooperatively with the grains surrounding them\cite{Heighway2019}, and thus sustain single slip. However, this mechanism does not extend to coarse-grained polycrystals, nor to single crystals; to reach the quasihydrostatic stress states typically observed in experiment, these specimens must generally deform on at least two slip systems. For this reason, we shall now consider the next-simplest case of idealized plastic deformation, namely equal slip on two complementary slip systems.

To treat double slip, one needs to generalize Eq.~(\ref{eq:Fp_otimes}) to the case of slip on $N$ slip systems. If we imagine that an incremental amount of slip occurs sequentially on slip systems labelled 1 to $N$, the resulting plastic deformation gradient reads
	\begin{equation} \label{eq:sequential_slip}
	F^p = (I + \gamma_N S_N)\ ...\ (1 + \gamma_2 S_2)(1 + \gamma_1 S_1),
	\end{equation}
where $\gamma_i$ is the amount of glide on slip system $i$ and $S_i = \mathbf{m}_i \otimes \mathbf{n}_i$ is its corresponding \emph{Schmid matrix}. While exact, this expression is inconvenient to work with due to the presence of its many cross-terms. For the purposes of developing a simple model, we shall take the leading-order approximation of Eq.~(\ref{eq:sequential_slip}), which reads
	\begin{equation} \label{eq:sequential_slip_linearised}
	F^p = I + \sum_{i=1}^N \gamma_i S_i.
	\end{equation}
The leading-order expression for the inverse of $F^p$ reads
	\begin{equation} \label{eq:sequential_slip_linearised_inverse}
	F^{p-1} = I - \sum_{i=1}^N \gamma_i S_i.
	\end{equation}
Note that in this approximation, the order of the plastic strains is immaterial. The ratio between the terms in Eq.~(\ref{eq:sequential_slip_linearised}) and the largest cross-terms it neglects scale with the typical glide $\tilde{\gamma}$, which, in a shock-compression context, will typically be of order 10\% or so. This figure gives an idea of the size of the error we introduce by using this linearized form of $F^p$ for our model.

\begin{figure}
\includegraphics{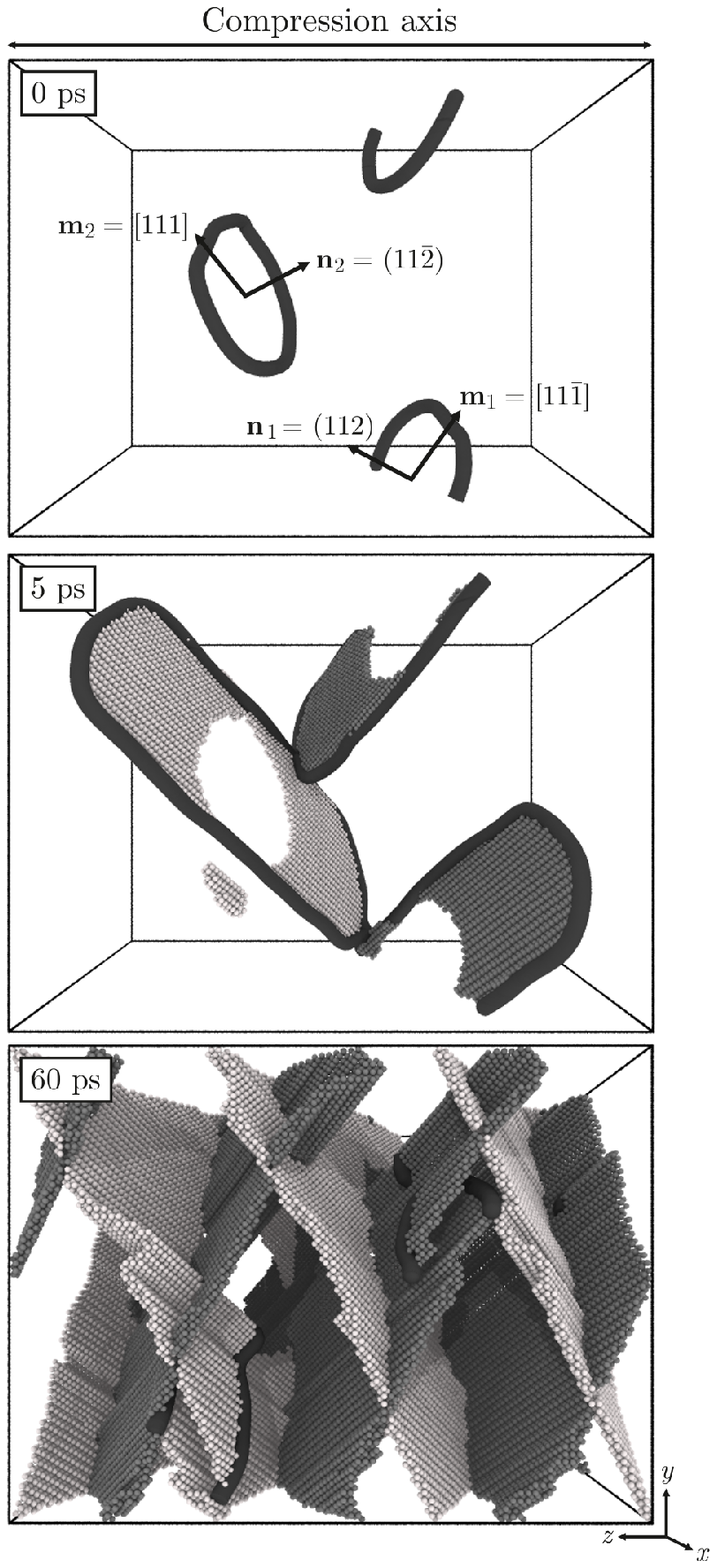}
\caption{\label{fig:double_slip_slideshow} Evolution of a fully periodic block of tantalum containing pre-existing dislocation loops of the types $[11\bar{1}](112)$ and $[111](11\bar{2})$ when compressed by 10\% along $[001]$. The crystal initially has dimensions of $19.9\times19.9\times29.8$\,nm\textsuperscript{3}, and has its $\langle100\rangle$ crystallographic directions aligned with the simulation cell axes. Shown are both the dislocation loops, which were characterized using the dislocation extraction algorithm (DXA) \cite{Stukowski2010b,Stukowski2012}, and the slipped atoms left in their wake as they propagate, which were identified using slip vector analysis (SVA)\cite{Zimmerman2001}. Atoms belonging to slip systems $[11\bar{1}](112)$ and $[111](11\bar{2})$ are colored dark and light grey, respectively. Atoms slipped during the injection of the loops have been omitted. Visualizations were performed using Ovito \cite{Stukowski2010}. }
\end{figure}

The simulation we use to investigate double slip differs from that described in the previous section in only two regards: the crystal is somewhat larger, with dimensions of $19.9\times19.9\times29.8$\,nm\textsuperscript{3}; and the crystal is injected with two different dislocations of the types $[11\bar{1}](112)$ and $[111](11\bar{2})$, which are deliberately offset from one another in the $y$ direction so that they can pass through one another freely. We show in Fig.~\ref{fig:double_slip_slideshow} the evolution of the dislocations loops and slip planes for the crystal deforming via conjugate slip. Much like the single-slip case, the dislocation loops rapidly expand along the direction of their respective Burgers vectors, and halt after around 30~ps when their corresponding slip planes uniformly fill the crystal. We observe again that the slip planes are rumpled in several planes, indicating a limited degree of cross-slip has taken place.

\begin{figure}
\includegraphics{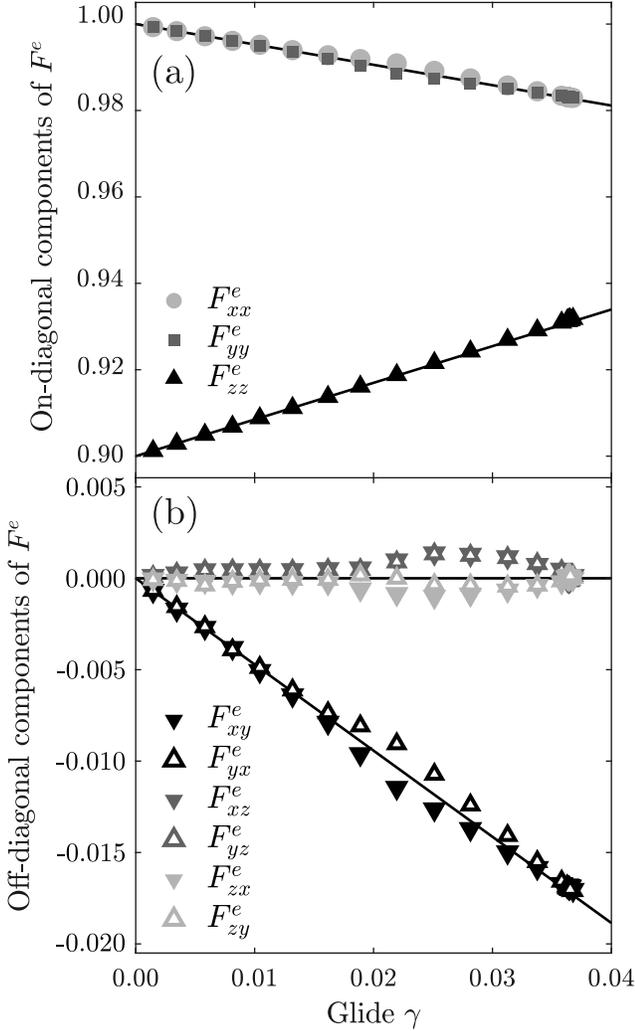}
\caption{\label{fig:loci_double} Evolution of the elements of the average elastic deformation gradient $F^e$ for a tantalum crystal uniaxially compressed by 10\% along $[001]$, deforming predominantly on its $[11\bar{1}](112)$ and $[111](11\bar{2})$ slip systems. Elements of $F^e$ are plotted as a function of the instantaneous glide $\gamma$ fitted from the elastoplastic decomposition. Solid lines are the loci to which the elements of $F^e$ should adhere. (a)~Evolution of the on-diagonal elements. (b)~Evolution of the off-diagonal elements.}
\end{figure}

\begin{figure}
\includegraphics{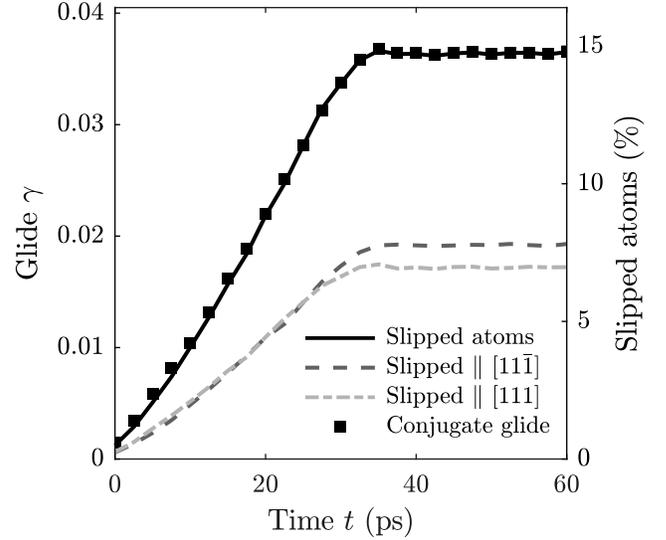}
\caption{\label{fig:glide_vs_slipped_atoms_double} Measurements of slip activity in a tantalum crystal uniaxially compressed along $[001]$, deforming predominantly on its $[11\bar{1}](112)$ and $[111](11\bar{2})$ slip systems. Discrete points show the glide derived from the elastic deformation gradient, while the curves show the number of atoms slipped along each Burgers vector, and the total number of slipped atoms.}
\end{figure}

For the present case of idealized conjugate slip, it may be shown using Eq.~(\ref{eq:sequential_slip_linearised_inverse}) that the locus of accessible elastic deformation states is this time given by
	\begin{equation} \label{eq:locus_double}
	F^e(\gamma) =
	\begin{pmatrix}
	1 - \frac{\sqrt{2}}{3}\gamma & -\frac{\sqrt{2}}{3}\gamma & 0 \\
	-\frac{\sqrt{2}}{3}\gamma & 1 - \frac{\sqrt{2}}{3}\gamma & 0 \\
	0 & 0 & \left(1 + \frac{2\sqrt{2}}{3}\gamma\right)v
	\end{pmatrix},
	\end{equation}
where $\gamma$ is the amount of glide on the two slip systems. In Fig.~\ref{fig:loci_double}, we compare once again the predictions of Eq.~(\ref{eq:locus_double}) with the measured variation in the elastic deformation gradient. Unlike in the single-slip case, we observe that the elements of $F^e$ depart slightly from their expected paths at intermediate times. This indicates the crystal is not deforming by perfect double slip, either because slip on one system slightly outpaces that on the other, or because the amount of cross slip alone is great enough to significantly change $F^e$. That being said, the agreement is convincing at early and late times, and, on the whole, the elements of $F^e$ follow their expected trajectories.

In Fig.~\ref{fig:glide_vs_slipped_atoms_double}, we compare the glide inferred from Eq.~(\ref{eq:locus_double}) with the number of slipped atoms detected using SVA. The correlation between these two independent measures of slip activity is very strong once again. As noted above, the slip is not exactly conjugate, at least not beyond 30~ps -- according to the SVA, there is 13\% more cumulative activity on slip systems with Burgers vector $[11\bar{1}]$ than on those with Burgers vector $[111]$. Our model functions fairly well in spite of this complication, and still allows us to infer the total slip activity from the evolution of this single crystal's unit cell.

\subsection{\label{sec:taylor_ambiguity} The Taylor ambiguity}

We have established that a simple model based upon the elastoplastic decomposition can be used to calculate the total amount of slip suffered by uniaxially strained crystals in the idealized cases of single and double slip. However, the success of the model was implicitly reliant on our knowing in advance which slip systems were operative. Suppose we were to approach these simulations `blind', without knowing \emph{a priori} which subset of the (say) $\langle111\rangle\{112\}$ slip systems would become active under compression. Is it possible simultaneously to identify the active slip systems and to deduce how much glide has taken place upon them?

Given a measurement of the elastic deformation gradient $F^e$, one can calculate the plastic deformation gradient directly using Eq.~(\ref{eq:decomposition}):
	\begin{subequations}
	\begin{align}
	F^p &= F^{e-1}F \\
	&= F^{e-1}\,\text{diag}(1,1,\text{det}F^e),
	\end{align}
	\end{subequations}
assuming as usual that the material is everywhere uniaxially strained. According to Eq.~(\ref{eq:sequential_slip_linearised}), this plastic deformation gradient is (to leading order) a point function of the glides $\{\gamma_i\}$ on $N$ independent slip systems:
	\begin{equation*}
	F^p = I + \sum_{i=1}^N \gamma_i S_i.
	\end{equation*}
If we consider only slip systems of the type $\langle111\rangle\{112\}$, there are 9 elements of $F^p$ (only 8 of which are independent, since $\text{det}\,F^p=1$) that are each a function of $N=12$ glides. The problem of identifying the operative slip systems is thus mathematically underconstrained. That is to say that there are infinitely many ways of combining slip on the 12 $\langle111\rangle\{112\}$ slip systems to yield the observed plastic strain state, and thus no way of uniquely determining the subset of slip systems responsible. This is the well-known \emph{Taylor ambiguity}\cite{Taylor1937, BishopandHill1951}. To arrive at a unique solution for the set of glides $\{\gamma_i\}$, one must use additional mathematical constraints derived from physical arguments about the mechanical properties of the material in question.

The solution we will test here is a variant of the \emph{minimum work principle}, first put forward by Taylor in the same paper that he proposed the full-constraints model\cite{Taylor1937}. Taylor argued that the crystal would deform in the least `wasteful' way possible, taking the pathway that dissipated the least amount of internal plastic work. In the special case that the flow stress of every slip system is identical (which, after work hardening begins, is by no means guaranteed), this postulate reduces to the \emph{minimum slip principle}. This states that the crystal reaches the final plastic strain state via the least amount of glide possible. Though highly simplistic, this argument is attractive because it is purely kinematic, and is therefore very easy to implement. We will show that while the minimum slip principle might in general represent an unwarranted oversimplification, it happens to be sufficient for the single- and double-slip cases considered here.

To implement the minimum slip principle, we first note that Eq.~(\ref{eq:sequential_slip_linearised}) relating $F^p$ to the glides $\{\gamma_i\}$, which reads
	\begin{equation}
	F^p=
	\begin{pmatrix}
	1 + \sum_i \gamma_i S_{i,xx} & \cdots & \sum_i \gamma_i S_{i,xz} \\
	\vdots & \ddots & \vdots \\
	\sum_i \gamma_i S_{i,zx} & \cdots & 1 + \sum_i \gamma_i S_{i,zz}
	\end{pmatrix},
	\end{equation}
may be recast into the form
	\begin{equation}\label{eq:unwrapped}
	\underbrace{\begin{pmatrix}
	E_{xx}^p \\
	\vdots \\
	E_{zz}^p
	\end{pmatrix}}_{\mathbf{E}^p}
	=
	\underbrace{\begin{pmatrix}
	S_{1,xx} & \cdots & S_{N,xx} \\
	\vdots & \ddots & \vdots \\
	S_{1,zz} & \cdots & S_{N,zz}
	\end{pmatrix}}_{S}
	\underbrace{\begin{pmatrix}
	\gamma_1 \\
	\vdots \\
	\gamma_N
	\end{pmatrix},}_{\boldsymbol{\gamma}}
	\end{equation}
where $\mathbf{E}^p$ is a 9-dimensional vector related to the plastic strain (whose components are defined such that $E_{ij}^p = F_{ij}^p - \delta_{ij}$), $S$ is an $N\times9$ matrix encoding the geometry of the slip systems, and $\boldsymbol{\gamma}$ is an $N$-dimensional vector containing the glides. The locus of solutions to Eq.~(\ref{eq:unwrapped}) can be expressed as \cite{James1978}
	\begin{equation}
	\boldsymbol{\gamma} = S^g\mathbf{E}^p + (I - S^g S)\mathbf{w},
	\end{equation}
where $S^g$ is the pseudoinverse of $S$, and the auxiliary $N$-dimensional vector $\mathbf{w}$ varies over all possible values. To find the solution that satisfies the minimum slip principle is simply a matter of locating the particular value of $\boldsymbol{\gamma}$  that minimizes the `cost function'
	\begin{equation}
	\Gamma(\boldsymbol{\gamma}) = \sum_{i=1}^N |\gamma_i|,
	\end{equation}
which we do here using a covariance matrix adaptation evolutionary strategy (CMA-ES).

\begin{figure}
\includegraphics{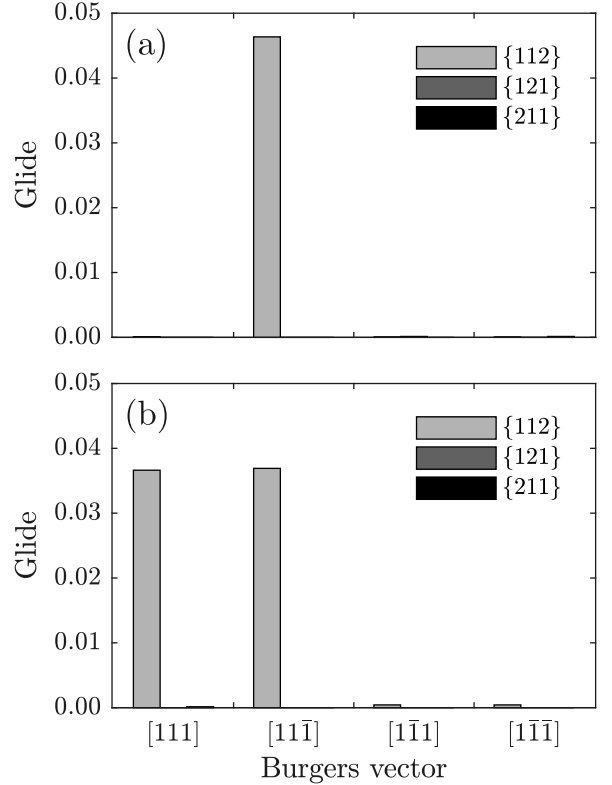}
\caption{\label{fig:bar_graphs} Calculated glides on the 12 $\langle111\rangle\{112\}$ slip systems in crystals containing (a) a single $[11\bar{1}](112)$ dislocation loop and (b) a $[11\bar{1}](112)$ loop and a $[111](11\bar{2})$ loop after being compressed by 10\% along $[001]$. Glides were calculated from the crystals' average elastic deformation gradient $F^e$, assuming this deformation state was reached via the least possible amount of total glide, summed over all the slip systems.}
\end{figure}

In Fig.~\ref{fig:bar_graphs}, we show the glides on the 12 $\langle111\rangle\{112\}$ slip systems inferred using the minimum slip criterion for the tantalum crystals undergoing single or double slip discussed in Secs.~\ref{sec:single_slip} and \ref{sec:double_slip}. The slip activity that the algorithm predicts is, on the whole, correct: in the single-slip case, we predict activity almost exclusively on the $[11\bar{1}](112)$ system, while glide in the double-slip case takes place overwhelmingly on the $[11\bar{1}](112)$ and $[111](11\bar{2})$ systems. Unsurprisingly, the prediction is not perfect: for the double-slipping crystal, for instance, we know from Fig.~\ref{fig:double_slip_slideshow} that some degree of cross-slip takes place, and might therefore hope to predict nonzero glides on not only the $[11\bar{1}](112)$ system, but also on $[11\bar{1}](\bar{1}21)$ and $[11\bar{1}](2\bar{1}1)$. This is not the case, however. It seems that the plastic strain effected by this cross-slip has instead been attributed to a small amount of activity on slip systems with $\mathbf{m} = [1\bar{1}1]$ or $[1\bar{1}\bar{1}]$, systems that the dislocation extraction algorithm (DXA)\cite{Stukowski2010b,Stukowski2012} reveals to be completely inactive. So, while we are able to recover the gross features of the crystals' plastic deformation modes, the price we pay for using the simplistic minimum slip principle is an inability to resolve their finer features, namely limited amounts of cross-slip. Use of the fully-fledged minimum work principle, accounting for the flow stress of each individual slip system, might allow one to capture these finer details.

In summary, when it is coupled to the minimum slip constraint, our simple kinematic model of plastic deformation under uniaxial strain conditions allows us both to identify the dominant subset of slip systems operative under compression and quantify their activity using only measurements of the elastic deformation gradient $F^e$, at least for the special cases of single and double slip.

\section{\label{sec:discussion} Discussion}

We have presented a simple kinematic framework based on the elastoplastic decomposition that we submit is better-suited to modelling the kind of uniaxially strained specimens encountered in a dynamic-compression context than are the usual Schmid and Taylor analyses. We have demonstrated how one might use this framework to measure slip activity in a uniaxially loaded monocrystal directly from measurements of its average elastic deformation $F^e$, the matrix that encodes the changes in size, shape, and orientation suffered by its unit cell. However, while we believe this \emph{in silico} study constitutes a small step in the right direction, more work will be required before a model such as this one can be applied with confidence to a real shock-compression experiment. In this section, we recapitulate the wider experimental context into which this model fits, and discuss which of its aspects warrant further scrutiny.

\begin{figure}
\includegraphics{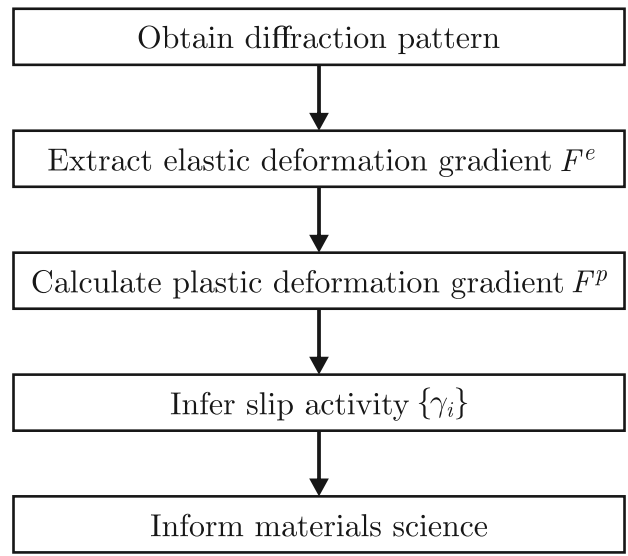}
\caption{\label{fig:workflow} Overview of the process by which experimental measurements of x-ray diffraction patterns may be used to study crystal plasticity under extreme conditions.}
\end{figure}

The ultimate ambition of a kinematic model like the one explored here is to enable one to gain an understanding of materials behavior under extreme loading conditions from experimental measurements of a sample's texture evolution. A schematic of the idealized workflow one might use to realize this ambition is shown in Fig.~\ref{fig:workflow}. Using an ultrabright x-ray source, one would first obtain time-resolved, \emph{in situ} diffraction images of a dynamically loaded crystalline sample. The angles at which the incoming x-rays are scattered depend on the separation and orientation of the specimen's atomic planes, and thus it is possible in principle to extract the elastic deformation gradient $F^e$ from the form of the diffraction pattern. With an appropriate kinematic model of the kind presented here, one can calculate the corresponding plastic deformation gradient $F^p$, from which one can (subject to the limits of the Taylor ambiguity) discern which slip systems are activated by the shock-compression process.

Steps one and two in this process are already well-developed: ultrafast x-ray diffraction is a well-developed diagnostic technique, and \emph{in situ} experimental measurement of the elastic deformation gradient\cite{Zaretsky2003} (or select components thereof, such as the rotation\cite{Turneaure2009, Suggit2012, Wehrenberg2017} or individual elastic strain components \cite{Turneaure2009, Murphy2010, Milathianaki2013, Comley2013, Wehrenberg2015, Sliwa2018}) obtained from diffraction data is becoming routine. The latter two steps, in which these elastic deformation gradients are converted into information about crystal plasticity, are relatively underdeveloped. As discussed in Sec.~\ref{sec:introduction}, in the two instances where analysis of this kind has been attempted in the past\cite{Wehrenberg2017, Suggit2012b, Suggit2012}, treatments based on the Schmid and Taylor analyses were used, which, for the reasons laid in Sec.~\ref{sec:rotation_rules}, we contend is not the correct approach. We believe the alternative model presented here gives a more faithful description of the kinematics of plasticity-induced texture evolution during planar compression. However, our model is incomplete, and we see three outstanding issues with it that require further investigation.

First, we have assumed throughout that the total deformation gradient $F$ is everywhere equal to the macroscopic deformation gradient to which the sample as a whole is subjected, namely $\text{diag}(1,1,v)$. While this assumption might be reasonable for single crystals, whether or not this should be true of a shock- or ramp-compressed \emph{polycrystal} is very much an open question. In general, one would expect the total deformation gradient $F$ to vary locally from grain to grain (and indeed within grains) due to the forces that each crystallite inevitably exerts upon its neighbors. The key question is whether the local deviations caused by such grain-grain interactions are `appreciable'. While there have in recent years been a few studies attempting to elucidate the nature of dynamic grain interactions under shock conditions \cite{Park2009, Luo2010, Heighway2019}, our understanding of this physics is far from complete. There is therefore, to the knowledge of the authors, no clear consensus about whether the full-constraints Taylor model used here is more valid than any other (such as the isostress model \cite{Sachs1928} or the self-consistent approach \cite{Kroner1961, Budiansky1962, Hill1965, Hutchinson1970, Lebensohn1993}) for loading to the megabar pressures of interest to the dynamic compression community. Consequently, there is no obvious way of knowing -- short of performing a spatially resolved and potentially very expensive simulation of the target in question -- whether the uniaxial form of $F$ is appropriate at the local level or not. There is a pressing need, then, for systematic investigations of polycrystals with a range of crystallographic and morphological textures undergoing dynamic compression, to ascertain when the Taylor model is adequate, and when finer adjustments to the mathematical form of $F$ are needed. In principle, these investigations could be supplemented by direct x-ray diffraction measurements: already, there exist models capable of predicting the expected form of the diffraction pattern from a textured sample in the limiting cases of uniform stress\cite{MacDonald2016} or elastic strain\cite{Higginbotham2014, McGonegle2015}. However, these models do not yet account for the plasticity-induced rotation suffered by each individual grain, limiting their applicability to cases in which such rotation is small.

Second, we have drawn attention to the fundamental limitation placed upon our approach by the Taylor ambiguity. This ambiguity means that if more than eight slip systems could conceivably be operative in a given target, one cannot uniquely determine the combination of glides $\{\gamma_i\}$ from purely kinematic considerations -- one must appeal to arguments about the material's mechanical properties to constrain the proportion of activity on each slip system. It so happened for the simple cases of single-  and double-slip considered here that the simplest possible physical constraint (the least slip principle) was sufficient to arrive at approximately the correct answer, but this will not always be the case, particularly if many more than two slip systems are active. Certain bcc metals, for instance, are liable to slip not only on their $\{112\}$ planes, but also on $\{110\}$ or $\{123\}$\cite{HullandBacon2011}, giving a total of 48 potentially active slip systems. In instances such as this where the glides $\{\gamma_i\}$ are hugely underconstrained, it is necessary to eliminate as many slip systems as possible from consideration (on the basis of their orientation relative to the compression axis, for example, or on their dislocation kinetics) before attempting to extract the glides on the operative systems. This procedure would call for a faithful strength model of the material in question, capable of capturing the strong pressure- and work-hardening effects that manifest under extreme loading conditions\cite{Barton2011}.

Third, we have restricted our attention here to plastic strain mediated by full dislocation slip. The situation in which plastic deformation is realized by the formation of stacking faults or deformation twins is more complicated, because there is no guarantee that the elastic deformation gradient $F^e$ describing the faulted material (defined with respect to its \emph{ideal} configuration) will be equal to that of the bulk material. That is to say that the elastic deformation gradient of the bulk material (which is what one directly measures from bulk diffraction peaks) differs from the volume-averaged elastic deformation gradient (which is what features in the elastoplastic decomposition). This subtle distinction could, if left unaccounted for, lead to spurious predictions of either the slip activity or the attending crystal rotation. Nonuniformity in $F^e$ caused by the creation of partial dislocations is another aspect of this model that would benefit from further dedicated computational study.

\section{\label{sec:conclusion} Conclusion}
We have shown using molecular dynamics simulations that the Schmid and Taylor treatments of plasticity-induced rotation are fundamentally unsuited to uniaxial dynamic compression conditions. We proposed an alternative treatment based on the full elastoplastic decomposition that we show correctly recovers the rotation and shear state of uniaxially strained single crystals undergoing single slip. We have further shown how such a framework may be used to infer the combination of slip systems responsible for the observed change in texture in the idealized cases of single and double slip. These results represent a modest but important step on the path towards a complete model with which one could deduce the slip activity of specimens dynamically loaded to extreme pressures via the evolution of their crystallographic texture.

\begin{acknowledgments}
The authors would like to thank P.~Avraam and D.~McGonegle for fruitful discussions, and M.~F.~Kasim for providing the implementation of the CMA-ES routine used here. Both P.~G.~H.~and J.~S.~W.~gratefully acknowledge the support of AWE via the Oxford Centre for High Energy Density Science (OxCHEDS), and J.~S.~W.~is further grateful to support from EPSRC under grant EP/S025065/1.
\end{acknowledgments}

\section*{Data availability}
The data that support the findings of this study are available from the corresponding author upon reasonable request.

\input{Kinematics_of_Rotation_V1.2.bbl}

\end{document}